\documentclass[final,5p,times,twocolumn]{elsarticle}
\usepackage{graphicx} 
\usepackage{amsmath} 

\newcommand{\br}{\mathbf{r}}

\newcommand{\bn}{\begin{equation}}
\newcommand{\ee}{\end{equation}}
\newcommand{\bga}{\begin{eqnarray}}
\newcommand{\eda}{\end{eqnarray}}

\newcommand{\diff}{\mathrm{d}}

\usepackage{amssymb}
\journal{journal for publication}

\begin{document}
\begin{frontmatter}

\title{van der Waals density functional calculations of binding in molecular crystals}
\author[a]{Kristian Berland}
\author[b]{\O yvind Borck}
\author[a]{Per Hyldgaard}
\address[a]{Department of Microtechnology and Nanoscience, MC2,
Chalmers University of Technology,
SE-41296 G\"{o}teborg, Sweden}
\address[b]{Department of Physics, Norwegian University of Science and Technology, H{\o}gskoleringen 5, NO-7491 Trondheim, Norway}

\begin{abstract}
A recent paper [J. Chem. Phys. {\bf 132} 134705 (2010)] illustrated the potential of the van der Waals density functional (vdW-DF) method [Phys. Rev. Lett. $\mathbf{92}$, 246401 (2004)] for efficient first-principle accounts of structure and cohesion in molecular crystals. 
Since then, modifications of the original vdW-DF version (identified as vdW-DF1) has been proposed, and there is also a new version called vdW-DF2 [ArXiv 1003.5255], within the vdW-DF framework. Here we investigate the performance and nature of the modifications 
and the new version for the binding of a set of simple molecular crystals:  hexamine, dodecahedrane, C60, and graphite. These extended systems provide benchmarks for computational methods dealing with sparse matter. 
We show that a previously documented  enhancement of non-local correlations of vdW-DF1 over an asymptotic atom-based  account close to and a few \AA\, beyond binding separation persists in vdW-DF2. The calculation and analysis of the binding in molecular crystals requires appropriate computational tools. In this paper, we also present details on our real-space parallel implementation of the vdW-DF correlation and on the method used to generate asymptotic atom-based pair potentials based on vdW-DF. 

\end{abstract}

\end{frontmatter}

\section{Introduction}

Supramolecular interactions such as steric hindrance, van der Waals (vdW) forces, and electrostatics play a central role in today's biological, nano-technology, and condensed-matter research. Materials where these interactions are important can be categorized as sparse matter \cite{sparseMatter}, since they have low electronic density in regions essential for cohesion and response. The all-pervasive vdW interactions act across such low-density regions. The need for robust computational tools that provide insight into supramolecular systems, have given impetus to the development of {\it first-principle} methods that properly accounts for the binding of sparse matter.

Density functional theory (DFT) within the local density approximation (LDA) or semi-local, generalized-gradient approximations (GGA) for the exchange-correlation potential  has, despite its tremendous success for dense matter, failed to consistently account for the binding in sparse matter. The van der Waals density functional (vdW-DF) method  --- both in the original version \cite{Dion:vdW,vdWDF:SC}, termed vdW-DF1, and in a new version, termed vdW-DF2 \cite{vdWDF2} --- go beyond these local approximations by using a non-local functional to approximate the correlation. 
Being first-principles, it deals with the vdW forces by including non-local correlations from the electron response to the electro-dynamical field. In the vdW-DF framework the non-local correlation energy takes the form of a double-space integral:
\bn
E_c^{\mathrm{nl}}[n] = \frac{1}{2} \int \diff \mathbf{r} \int d\mathbf{r}' 
n(\mathbf{r}) \,
\phi(\mathbf{r},\mathbf{r}') 
n(\mathbf{r}')\,.
\label{eq:Ecnl}
\ee
The vdW-DF method inherits the excellent performance of GGAs for many dense matter systems, while extending the reach of DFT approximations to sparse matter.

\begin{figure}[t]
\centering
\includegraphics[width=7cm]{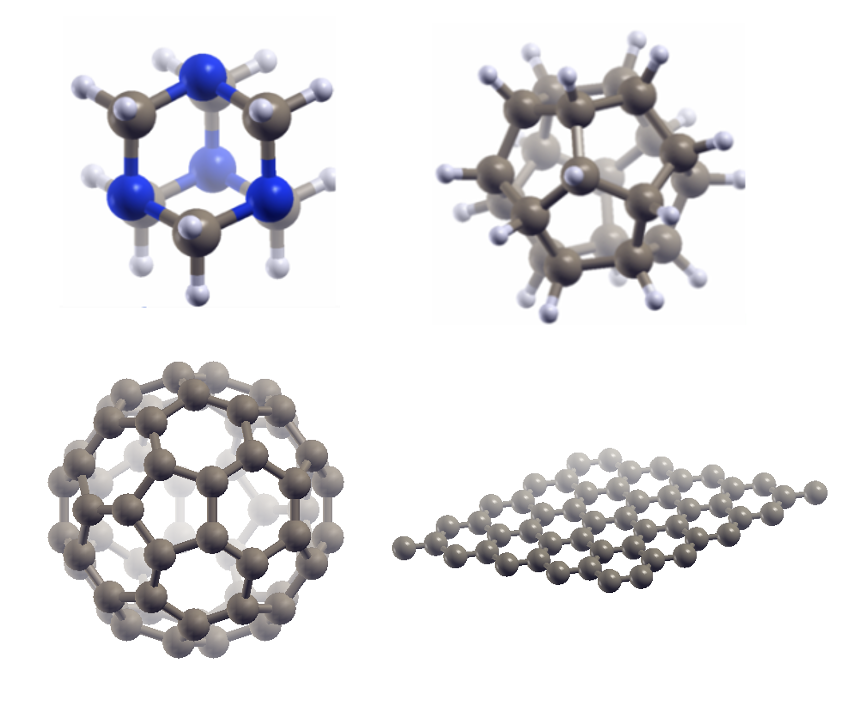}
\label{Fig:mol}
\caption{Molecules forming simple model crystals bound by van der Waals forces: hexamine, dodecahedrane, buckmeisterfullerene/C60, and graphene. (from top left to bottom right).}
\end{figure}

The original version of vdW-DF1 has performed well for a range of systems, such as binding in molecular dimers, physisorption on surfaces, and polymer crystals \cite{sparseMatter}. Two of us have shown that it accounts for the structure of molecular crystals \cite{molcrys}. However, vdW-DF1 \cite{Dion:vdW} overestimates binding separations by $0.2$ to $0.3\, \AA$. This has led several authors to suggest adjustments in the vdW-DF method. Using the S22 \cite{S22} data set of molecular dimers, Klime\v{s} {\it et al} \cite{Klimes} optimized several GGA exchange flavors for use in place of the original choice of revPBE$_{\rm x}$ \cite{revPBE}  (the x subscript indicates the exchange-part of the exchange-correlation) to the correlation in vdW-DF1; here we test optPBE$_{\rm x}$. Cooper developed an exchange flavor (C09$_{\rm x}$) for vdW-DF1, \cite{exchange:Cooper} which goes like the gradient-expansion approximation (GEA) in the slowly-varying, high-density, limit.   Very recently, a new version vdW-DF2 was proposed \cite{vdWDF2}. It modifies the inner functional, that vdW-DF uses to determine the local plasmon frequency $\sim q_0(\br)^2$ and thereby the non-local correlation. vdW-DF2 also uses a refitted version of the PW86$_{\rm x}$ \cite{PW86}. This exchange functional does not induce unphysical binding effects in the exchange channel and is simultaneously less repulsive than revPBE$_{\rm x}$ \cite{Langreth:exchange}. 

Computational methods designed for sparse matter benefit from using molecular crystals as testing grounds, because of the wealth of accurate experimental data on crystal structures and lattice parameters. Molecular crystals are also model system for bulk sparse matter. We therefore find it pressing to test vdW-DF2 and the suggested modifications of vdW-DF1 on these systems. Here, we extend our vdW-DF1 study of hexamine and dodecahedrane \cite{molcrys} with these new modifications, and include two important carbon allotropes: the C60 crystal and graphite. The latter can be viewed as a molecular crystal of graphene flakes. For the vdW-DF correlations (termed vdW-DF1$_c$ and vdW-DF2$_c$), we find that that both vdW-DF1$_c$ and vdW-DF2$_c$ are considerably enhanced over a corresponding asymptotic atom-based pair potential form at, and a few \AA\, beyond, typical binding separations. 

The vdW-DF calculations for molecular crystals require an implementation that handle periodic systems; in particular, for the C60 crystal, having four molecules (240 carbon atoms) per unit cell, a moderately efficient and parallel implementation is beneficial.
In this paper, we discuss some details of our parallel  code for evaluating non-local correlations, in addition to discussing the method used to generate results in the asymptotic approximation ('app').  

\section{Computational methods}

\subsection{van der Waals density functional calculations}

In the vdW-DF framework, the exchange-correlation of consists of LDA correlation, GGA exchange, and the non-local correlations:
\bn
E^{\rm vdW-DF}_{\rm xc}=E_c^{\rm LDA}+E_x^{\rm GGA} + E_c^{\mathrm{nl}}[n]\,.
\ee
The total energy, $E^{\rm vdW-DF}$, is obtained as in Refs. \cite{molcrys,benzCu,Potassium}.
The exchange-correlation energy is evaluated non-self consistently using the charge density, obtained in a self-consistent DFT calculation with the PBE \cite{PBE} flavor of GGA (sc-GGA), utilizing the ultrasoft pseudo-potential plane-wave code {\verb DACAPO } \cite{DACAPO}. 

The potential energy of the crystal is the difference between the total energy in a configuration and the energy of isolated molecules: $
E(a)= E^{\rm vdW-DF }(a)- E^{\rm vdW-DF}(a\rightarrow \infty) \, . 
$
We use brute force to determine the optimal value of the unit cell dimension (denoted $a$). The molecular structure, obtained in a sc-GGA calculation, is kept constant as $a$ is varied. The experimental crystal symmetry specifies the molecular orientations \cite{molcrys} for all but the C60 crystal, where we use the experimental orientations \cite{C60:nature}.

\subsection{Implementation of the non-local correlation integral}
 
A parallel implementation of the non-local correlation within the vdW-DF framework used in recent applications \cite{benzCu,molcrys,VO5} is described here. It evaluates the non-local correlation of Eq.~(\ref{eq:Ecnl}) by using an input charge grid provided by a software package. Our approach shares this post-processing strategy, using code-independent post-GGA evaluation of
the vdW-DF method, with the {\verb JuNoLo  \cite{JuNoLo} code.

\begin{figure}[t]
\centering
\includegraphics[width=5cm]{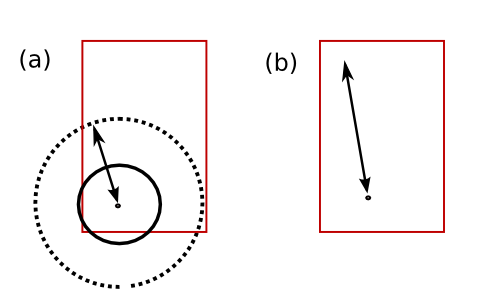}
\label{Fig:int}
\caption{Schematics of the evaluation the six-dimensional integral with (a) and without (b) a radius cutoff. The box shows the unit cell, the arrow connect to coordinates, the filled (dashed) circle indicates a radius cutoff.} 
\end{figure}

Figure \ref{Fig:int} illustrates the evaluation of the six-dimensional integral in real space using multiple radius cutoffs: $R_1$ and $R_2$ \cite{molcrys,Prl:svetla,JesperPhD}.
The inner domain defined by $|\br-\br'|<R_1$ is sampled on a dense grid, while the outer, defined by $R_1<|\br-\br'|<R_2$, is sampled on a grid with half the grid-point density. 
Since the integral is six-dimensional, the cpu cost is 64 times smaller \cite{Potassium} per volume in the inner domain than in the outer.
For periodic systems or systems with dimensions much larger than these radiuses, the cpu cost scales linearly with number of grid points $N$, but with a large prefactor $\sim R^3$.  For tiny, non-periodic system, the entire volume of the system falls within the domain and the scaling remains $N^2$. However, depending on the grid sampling density, a small inner cutoff $R_1$ can be used, and the costly part of the integral usually scales with $N$.

In discrete form, the six-dimensional integral of Eq.~(\ref{eq:Ecnl}) can be written as a double sum:
$E^{\rm nl}_{\rm c}=~\left(\Delta V\right)^2 \sum_i\sum_j n(\br_i) \phi_{ij}n(\br_j)\,,$
where the indices $i,j$ go over the entire grid. 
To evaluate a sum in parallel, different parts of the sum are distributed on the $M$ processors. A balanced load is achieved by splitting the sum according to summations in integer steps:
\bn
\sum_i = \underbrace{\sum_{Mi}}_{\rm proc.\, 1} +  \underbrace{\sum_{Mi+1}}_{\rm proc.\, 2}+ \underbrace{ \sum_{Mi+2}}_{\rm proc.\, 3} + ... + \underbrace{\sum_{M(i+1)-1}}_{\mathrm{proc.}\,  M}\,.
\ee

A scheme using interpolation to express Eq.~(\ref{eq:Ecnl})  in terms of convolutions and  achieves $\sim N\log N$ scaling has recently been reported \cite{Soler:speedup}. 
This scheme provides a significant speedup for medium-size systems. However, the real-space version has certain merits. First, the total cpu cost remains nearly constant for any number of processors. Second, for large systems the linear scaling in system size $N$ will be important. Third, the explicit control over the integration domain is useful both in analysis of binding and for systems which are nonperiodic in some or all spatial directions; for instance, the tertiary structure of biopolymers are characteristically non-periodic. The explicit control over length scales, can also be used together with coarser accounts of the vdW forces at large separations \cite{Kleis:Nanotube,Kleis:polymer1,JesperPhD}.

\subsection{Generating asymptotic vdW-DF pair potentials using Bader analysis}

The asymptotic atom-based pair potential approximation for the vdW energy between two molecules takes the form
\begin{equation}
E^{\rm vdW-DF}_{\rm app} = \sum_{i} \sum_j \frac{C_6^{ij}}{|\br_i-\br'_j|}\,, \label{eq:app}
\end{equation}
where $i$ and $j$ are labels for the atoms in of the two molecules. 
To generate such a potential based on vdW-DF, we partition the charge density of the molecule among the atoms, using the atoms-In-Molecules (AIM) method of Bader \cite{baderbook}, and evaluate the $C^{ij}_6$ coefficients for the different atomic pairs \cite{JesperPhD}. This section details the asymptotic form of vdW-DF, the AIM method, and its implementation. 

The $C_6$ coefficient for the asymptotic vdW interaction between two fragments, $A$ and $B$, takes the general form:
\bn
C_6^{AB}= \frac{3}{\pi}\int_0^\infty \diff u \, \alpha_A(iu) \alpha_B(iu) \,. \label{eq:C6}
\ee
In vdW-DF the polarizability of the fragment,
\bn
\alpha^{\rm vdW-DF}_A(\omega)= \int \diff^3 \br \,\chi^{\rm vdW-DF}_A(\omega, \br)\,, \label{eq:pol}
\ee	
is obtained by integrating over the local susceptibility:
\bn
\chi^{\rm vdW-DF}_A(\omega, \br) = \frac{n_A(\br)}{\left[9q_0(\br)^2/8\pi \right]^2 -\omega^2}\,,
\ee
This form is valid both for vdW-DF1 and for vdW-DF2, since they only differ in how they determine the inner functional specifying the local plasmon frequency $\sim~q_0(\br)^2$ \cite{Dion:vdW,vdWDF:SC,vdWDF2}. 

AIM partitions the the total charge-density of a molecule or a solid into atomic volumes $\Omega$
based solely on the topological properties of the charge-density.
The surfaces separating the atomic volumes are defined by $ \nabla n(\mathrm{r})\cdot {\bf l}=0$
where $\bf l$ is the unit vector normal to the surface. Atomic properties,
like charge and magnetic moments \cite{obes}, are obtained by integrating over the individual volumes. 
We use this partition to obtain the polarizability of an atom~$A$ using Eq.~(\ref{eq:pol}), with $n_A(\br) = \int_{\Omega_\mathrm{A}} \diff^3 \, \br' n(\br') \delta(\br-\br')$. 
The $q_0(\br)^2$ grid is generated prior to this partition to avoid unphysical gradients in the boundary regions. 

We use and extend a code implementation \cite{Oyvind} of the algorithm proposed by Henkelman {\it et al} \cite{henkelman}, to generate Bader volumes and determine the atoms-in-molecule polarizabilities of Eq.~(\ref{eq:pol}}).  As the pseudo-potential calculations generate a pseudo-electron density, the first step is to include the core-electron density to obtain the total density. The core-electron densities are generated upon producing the pseudopotentials. For each grid point, a path of steepest descent is constructed. The set of paths 
terminating at the same maximum of the charge density, are assigned to the same 
Bader volume. 

The paths are constructed as in Ref. \cite{henkelman}: First, the charge density of a given grid point is compared that of all 26 adjacent grid points. If it is larger than its neighbors, it is considered a maximum; if not, the algorithm proceeds to the adjacent grid point with the largest charge density. This procedure is repeated until a local maximum is reached. It also terminates if it reaches a grid point that belongs to a previously assigned Bader volume. The algorithm scales
linearly with $N$, because it only make a single loop for each grid-point.

\section{Crystal structure}

The molecular crystals of this study have simple structures. Hexamine and dodecahedrane forms respectively a body and face-centered cubic (bcc, fcc) with a single molecule per unit cell \cite{molcrys}. At low temperatures, the C60 crystal is a simple cubic, with four molecule per cell. At higher temperatures, the molecules rotate freely and the crystal becomes effectively an fcc \cite{C60:nature}. Graphite is a layered material, with two graphene sheets stacked in an AB pattern. 

\begin{figure}[h]
\centering
\includegraphics[width=7cm]{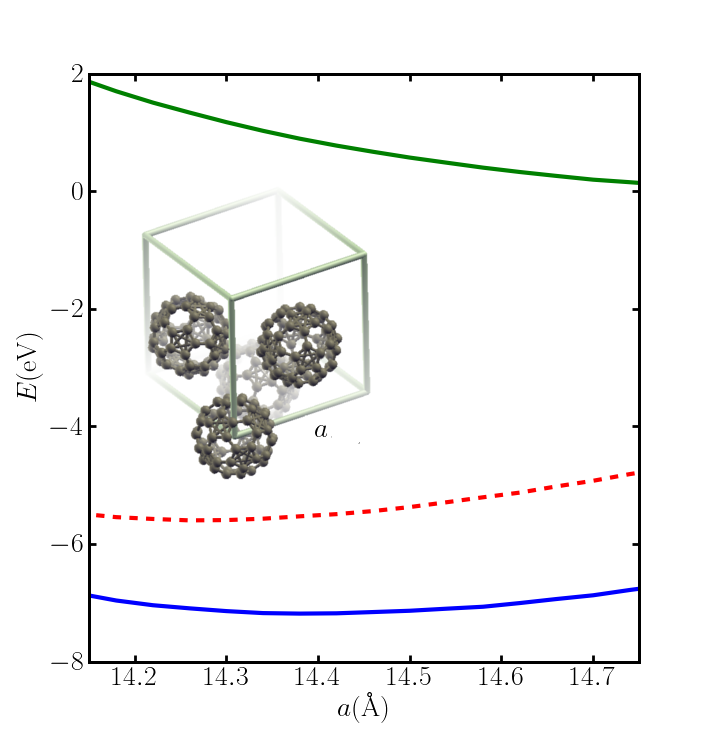}
\label{Fig:C60curve}
\caption{The binding curve of the C60 crystal. The lower [upper] (dashed) curve gives the vdW-DF1 [PBE] (vdW-DF2) result. The insert shows the simple cubic unit cell of the crystal, which becomes face-centered cubic only if internal orientation are neglected.}
\end{figure}

Figure \ref{Fig:C60curve} shows the binding curve of the C60 crystal as a function of the lattice parameter $a$ of the simple cubic. For this crystal, vdW-DF1 binds stronger than vdW-DF2, but at a larger lattice constant. The failure of DFT within GGA illustrate the need for non-local correlations to account for the binding in these crystals. 

\begin{table*}[t]
\label{tab:res}
\caption{vdW-DF predictions of lattice parameters and cohesion energy the crystals of hexamine, dodecahedrane, C60, and graphite, compared with experimental values, for different combination of exchange and correlation. The experimental lattice parameters are based on low temperature measurements, except for dodecahedrane. The cohesion energies of the cage crystals (graphite) are given per molecule (atom). The bold letters identifies the results for vdW-DF1 and vdW-DF2 versions.}

\begin{tabular}{lllll}
\hline
Functional / molecule: & {\it Hexamine} & {\it Dodecahedrane} & {\it Buckyball (C60)}  
& {\it Graphite}   \\
\begin{tabular}{ll}
Correlation &Exchange  \vspace{0.2cm}\\

{\bf vdW-DF1}$_{\rm c}$ &{\bf revPBE}$_{\rm x}$  \\
&C09$_{\rm x}$	\\
 &optPBE$_{\rm x}$ 	\vspace{0.2cm}\\

 {\bf vdW-DF2}$_{\rm c}$  &{\bf PW86}$_{\rm x}$\\
 Exp. &  
\end{tabular}
&
\begin{tabular}{ll}
$a$ (\AA)   &$E_{coh}$(eV)  \vspace{0.2cm} \\
  {\bf 7.16} & {\bf 1.01}  \\
     6.92     &  1.42 \\    
 6.96      &  1.21	\vspace{0.2cm}   \\

 {\bf 6.96} & {\bf 0.93}  \\
    
 6.91 \footnotemark[1]     & 0.83 \footnotemark[2] 
\end{tabular}
&
\begin{tabular}{ll}
  $a$ (\AA)   &$E_{coh}$(eV)  \vspace{0.2cm}\\
  {\bf 10.92 } & {\bf 1.46} \\
 10.44 & 1.65 \\
 10.64 & 1.75\vspace{0.2cm}  \\

 {\bf 10.64} & {\bf 1.35 } \\

  10.60\footnotemark[3] & -
\end{tabular}
&
\begin{tabular}{ll}
  $a$ (\AA)   &$E_{coh}$(eV)  \vspace{0.2cm}\\
   {\bf 14.38} & {\bf1.70} \\
  14.10  &  1.98\\
14.22 & 2.02 \vspace{0.2cm} \\

{\bf 14.30}  & {\bf 1.30}\\
 14.04 \footnotemark[4] &   1.6-1.9 \footnotemark[5] 

\end{tabular}
&
\begin{tabular}{ll}
  $c$ (\AA)   &$E_{coh}$(eV)  \vspace{0.2cm}\\
{\bf  7.24}   & {\bf0.053}\\
6.44 &0.073 \\
\vspace{0.2cm} 
 6.88 &0.064 \\
{\bf 6.96} & {\bf 0.053}\\

6.67\footnotemark[6] & 0.052\footnotemark[7]
\end{tabular} 
\\ \hline
\end{tabular}

\footnotesize{
\footnotemark[1]{Ref.~\cite{Hexamine:Struc}},
\footnotemark[2]{Ref.~\cite{Hexamine:Cohesion}},
\footnotemark[3]{Ref.~\cite{dod:struc}},
\footnotemark[5]{Ref.~\cite{chemref}},
\footnotemark[4]{Ref.~\cite{C60:nature}},
\footnotemark[6]{Ref.~\cite{graphite:struc}},
\footnotemark[7]{Ref.~\cite{coh:graphite}},

}.
\label{table:Crystal}
\end{table*}

Table \ref{tab:res} shows the lattice constants and cohesion energies obtained using the vdW-DF method. The lattice constants predicted by vdW-DF1 overestimates the experimental values by about 0.3~\AA~(per sheet for graphite), which is consistent with earlier studies \cite{sparseMatter}. The cohesion energies compares well with experimental values. The use of C09$_{\rm x}$ and optPBE$_{\rm x}$ as exchange partner to the vdW-DF1$_c$ improves lattice constants. The former is almost spot-on, but slightly underestimates them, while the latter overestimates them. Since the experimental value for Dodecaherane is based on room-temperature results, C09$_{\rm x}$ compares well with experiments also for this molecule. In both these modifications of vdW-DF1, the predicted cohesion energies are quite large compared to the experimental ones. vdW-DF2 improves the lattice constants over vdW-DF1, giving an overestimation similar to optPBE$_{\rm x}$, while the cohesion energies are somewhat reduced.

vdW-DF2 has the the best overall performance. The calculated vdW-DF2 cohesion energy for the C60 crystal is lower than experimental observations. For this crystal, internal orientations are not optimized and this might contribute to the increased difference with the experimental data for all the modifications. The C60 molecule also differs from the others by having a low surface to volume ratio.

\section{Enhancement over asymptotic pair-potential form}

Previous studies \cite{molcrys,Kleis:polymer1,Kleis:Nanotube} have shown that the non-local correlation of vdW-DF1 is enhanced compared to an atomic-based asymptotic pair-potential ('app') account (Eq.~\ref{eq:app}) at binding separation and a few \AA\, beyond. By construction 'app' does not include the image-plane and multi-pole effects inherent in the density-functional framework of vdW-DF. The importance of image planes were also discussed in the seminal work of Zaremba and Kohn in a surface-physics context \cite{ZarembaKohn}.  Asymptotic atom-based approximations are often used in force-field methods and in semi-empirical methods that add these interactions on top of GGA based DFT calculations (DFT-D). Since vdW-DF2 modifies the account of non-local correlation, we here investigate how this affects the enhancement over 'app'. 

\begin{figure}[h]
\includegraphics[width=8cm]{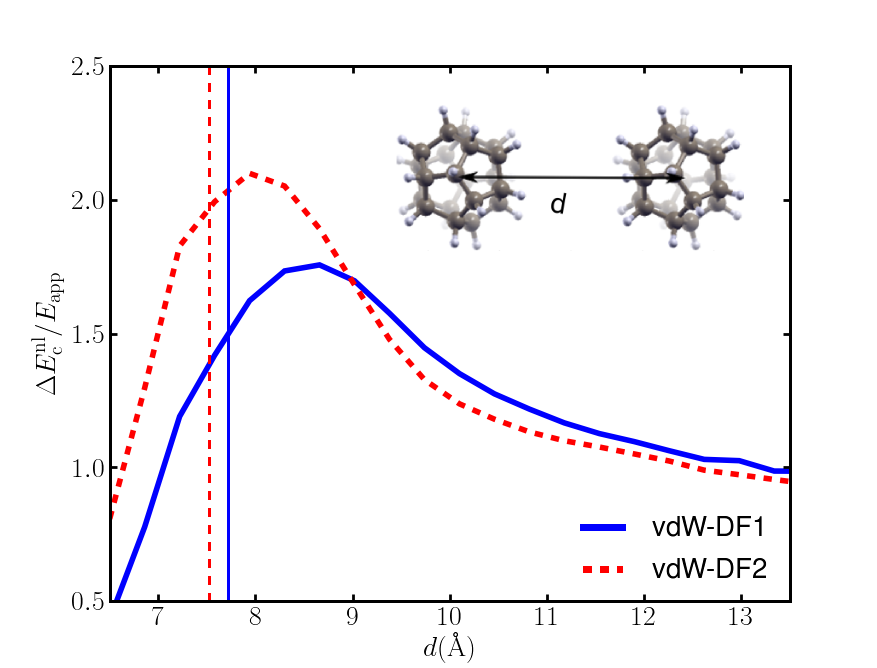} 
\caption{Role of of image-plane and multi-pole effects in the binding of a dodecahedrane dimer: The enhancement of non-local correlation of vdW-DF (1 and 2) over the corresponding asymptotic atom-based pair-potential form. 
The full (dashed) line give the result for vdW-DF1 (vdW-DF2). The vertical lines indicate the nearest-neighbor binding separation in the dodecahedrane crystal.}
\label{fig:domdimer}
\end{figure}

Figure \ref{fig:domdimer} shows the ratio between the non-local correlation and its corresponding 'app', $\Delta E^{nl}_{c}/E_{\rm app}$, for both vdW-DF1 and vdW-DF2. 
Because of residual noise in our determination of the $C_6$ coefficients, we have ($<10~\%$) adjusted the curves to reach unity at $d>16 \AA$.
 The full (dashed) line gives the result for vdW-DF1 (vdW-DF2). 
Both version exhibit a significant enhancement, with the vdW-DF2 result being somewhat larger and shifted to smaller separations. The less than unity ratio at separations about $1 {\rm \AA}$ shorter than the binding separations reflects the built-in damping of the vdW forces in the vdW-DF framework; no  ad hoc damping parameter needs to be introduced in the vdW-DF method.

\section{Discussion and summary}
The performance of variations of the vdW-DF method has been investigated for several molecular crystals. The modifications of vdW-DF1, involving only the exchange functional (C09$_{\rm x}$, optPBE$_{\rm x}$), improves the lattice constants, but cohesion energies seem to worsen for molecular crystals. That optPBE$_{\rm x}$ did not outperform the other vdW-DF1 combinations indicate that fitting to S22, by itself, does not guarantee high precision for other types of systems. C09$_{\rm x}$ gives excellent lattice parameters. The good performance of vdW-DF2 is encouraging, pointing to higher accuracy for the vdW-DF method. 

The enhancement of non-local correlation over 'app' shows that the nature of binding both for vdW-DF1 and vdW-DF2 differs from the asymptotic 'app' description. This result further questions the often used $1/r^6$ atom-based approximation for van der Waals forces when used within a few \AA\, of the binding separations \cite{molcrys,Kleis:Nanotube}.

\section*{Acknowledgment}
We thank E. Londero, J. Rohrer, and E. Schr\"oder for valuable discussions and sharing code and numerical tests. J. Kleis and E. Schr\"oder are acknowledged for implementing an earlier serial version of the code described here (together with PH). The Swedish National Infrastructure for Computing (SNIC) is acknowledged for allocation of computer time and for supporting KB’s participation in the National Graduate School in Scientific Computing (NGSSC). The work was supported by the  Swedish research  Council (Vetenskapsr\aa det VR) under 621-2008-4346. 


\begin{thebibliography}{34}
\expandafter\ifx\csname natexlab\endcsname\relax\def\natexlab#1{#1}\fi
\providecommand{\bibinfo}[2]{#2}
\ifx\xfnm\relax \def\xfnm[#1]{\unskip,\space#1}\fi
\bibitem[{Langreth et~al.(2009)Langreth, Lundqvist, Chakarova-K\"ack, Cooper,
  Dion, Hyldgaard, Kelkkanen, Kleis, Kong, Li, Moses, Murray, Puzder, Rydberg,
  Schr\"oder, and Thonhauser}]{sparseMatter}
\bibinfo{author}{D.~C. Langreth}, \bibinfo{author}{B.~I. Lundqvist},
  \bibinfo{author}{S.~D. Chakarova-K\"ack}, \bibinfo{author}{V.~R. Cooper},
  \bibinfo{author}{M.~Dion}, \bibinfo{author}{P.~Hyldgaard},
  \bibinfo{author}{A.~Kelkkanen}, \bibinfo{author}{J.~Kleis},
  \bibinfo{author}{L.~Kong}, \bibinfo{author}{S.~Li}, \bibinfo{author}{P.~G.
  Moses}, \bibinfo{author}{E.~Murray}, \bibinfo{author}{A.~Puzder},
  \bibinfo{author}{H.~Rydberg}, \bibinfo{author}{E.~Schr\"oder},
  \bibinfo{author}{T.~Thonhauser},
\newblock \bibinfo{title}{A density functional for sparse matter},
\newblock \bibinfo{journal}{Journal of Physics: Condensed Matter}
  \bibinfo{volume}{21} (\bibinfo{year}{2009}) \bibinfo{pages}{084203}.
\bibitem[{Dion et~al.(2004)Dion, Rydberg, Schr\"oder, Langreth, and
  Lundqvist}]{Dion:vdW}
\bibinfo{author}{M.~Dion}, \bibinfo{author}{H.~Rydberg},
  \bibinfo{author}{E.~Schr\"oder}, \bibinfo{author}{D.~C. Langreth},
  \bibinfo{author}{B.~I. Lundqvist},
\newblock \bibinfo{title}{Van der waals density functional for general
  geometries},
\newblock \bibinfo{journal}{Phys. Rev. Lett.} \bibinfo{volume}{92}
  (\bibinfo{year}{2004}) \bibinfo{pages}{246401}.
\bibitem[{Thonhauser et~al.(2007)Thonhauser, Cooper, Li, Puzder, Hyldgaard, and
  Langreth}]{vdWDF:SC}
\bibinfo{author}{T.~Thonhauser}, \bibinfo{author}{V.~R. Cooper},
  \bibinfo{author}{S.~Li}, \bibinfo{author}{A.~Puzder},
  \bibinfo{author}{P.~Hyldgaard}, \bibinfo{author}{D.~C. Langreth},
\newblock \bibinfo{title}{Van der waals density functional: Self-consistent
  potential and the nature of the van der waals bond},
\newblock \bibinfo{journal}{Phys. Rev. B} \bibinfo{volume}{76}
  (\bibinfo{year}{2007}) \bibinfo{pages}{125112}.
\bibitem[{{Lee} et~al.(2010){Lee}, {Murray}, {Kong}, {Lundqvist}, and
  {Langreth}}]{vdWDF2}
\bibinfo{author}{K.~{Lee}}, \bibinfo{author}{{\'E}.~D. {Murray}},
  \bibinfo{author}{L.~{Kong}}, \bibinfo{author}{B.~I. {Lundqvist}},
  \bibinfo{author}{D.~C. {Langreth}},
\newblock \bibinfo{title}{{A Higher-Accuracy van der Waals Density
  Functional}},
\newblock \bibinfo{journal}{ArXiv e-prints}  (\bibinfo{year}{2010}).
\bibitem[{Berland and Hyldgaard(2010)}]{molcrys}
\bibinfo{author}{K.~Berland}, \bibinfo{author}{P.~Hyldgaard},
\newblock \bibinfo{title}{Structure and binding in crystals of cagelike
  molecules: Hexamine and platonic hydrocarbons},
\newblock \bibinfo{journal}{The Journal of Chemical Physics}
  \bibinfo{volume}{132} (\bibinfo{year}{2010}) \bibinfo{pages}{134705}.
\bibitem[{Jure\v{c}ka et~al.(2006)Jure\v{c}ka, \v{S}poner, \v{C}erny, and
  Hobza}]{S22}
\bibinfo{author}{P.~Jure\v{c}ka}, \bibinfo{author}{J.~\v{S}poner},
  \bibinfo{author}{J.~\v{C}erny}, \bibinfo{author}{P.~Hobza},
\newblock \bibinfo{title}{Benchmark database of accurate (mp2 and ccsd(t)
  complete basis set limit) interaction energies of small model complexes, dna
  base pairs, and amino acid pairs},
\newblock \bibinfo{journal}{Physical Chemistry Chemical Physics}
  \bibinfo{volume}{8} (\bibinfo{year}{2006}) \bibinfo{pages}{1985--1993}.
\bibitem[{Klime\v{s} et~al.(2010)Klime\v{s}, Bowler, and Michaelides}]{Klimes}
\bibinfo{author}{J.~Klime\v{s}}, \bibinfo{author}{D.~R. Bowler},
  \bibinfo{author}{A.~Michaelides},
\newblock \bibinfo{title}{Chemical accuracy for the van der waals density
  functional},
\newblock \bibinfo{journal}{Journal of Physics: Condensed Matter}
  \bibinfo{volume}{22} (\bibinfo{year}{2010}) \bibinfo{pages}{022201}.
\bibitem[{Zhang and Yang(1998)}]{revPBE}
\bibinfo{author}{Y.~Zhang}, \bibinfo{author}{W.~Yang},
\newblock \bibinfo{title}{Comment on ``{Generalized Gradient Approximation Made
  Simple}''},
\newblock \bibinfo{journal}{Phys. Rev. Lett.} \bibinfo{volume}{80}
  (\bibinfo{year}{1998}) \bibinfo{pages}{890}.
\bibitem[{{Cooper}(2010)}]{exchange:Cooper}
\bibinfo{author}{V.~R. {Cooper}},
\newblock \bibinfo{title}{van der waals density functional: An appropriate
  exchange functional} \bibinfo{volume}{81} (\bibinfo{year}{2010})
  \bibinfo{pages}{161104}.
\bibitem[{Perdew and Yue(1986)}]{PW86}
\bibinfo{author}{J.~P. Perdew}, \bibinfo{author}{W.~Yue},
\newblock \bibinfo{title}{Accurate and simple density functional for the
  electronic exchange energy: Generalized gradient approximation},
\newblock \bibinfo{journal}{Phys. Rev. B} \bibinfo{volume}{33}
  (\bibinfo{year}{1986}) \bibinfo{pages}{8800--8802}.
\bibitem[{Murray et~al.(2009)Murray, Lee, and Langreth}]{Langreth:exchange}
\bibinfo{author}{E.~D. Murray}, \bibinfo{author}{K.~Lee},
  \bibinfo{author}{D.~C. Langreth},
\newblock \bibinfo{title}{Investigation of exchange energy density functional
  accuracy for interacting molecules},
\newblock \bibinfo{journal}{Journal of Chemical Theory and Computation}
  \bibinfo{volume}{5} (\bibinfo{year}{2009}) \bibinfo{pages}{2754--2762}.
\bibitem[{Berland et~al.(2009)Berland, Einstein, and Hyldgaard}]{benzCu}
\bibinfo{author}{K.~Berland}, \bibinfo{author}{T.~L. Einstein},
  \bibinfo{author}{P.~Hyldgaard},
\newblock \bibinfo{title}{Rings sliding on a honeycomb network: Adsorption
  contours, interactions, and assembly of benzene on {Cu(111)}},
\newblock \bibinfo{journal}{Phys. Rev. B} \bibinfo{volume}{80}
  (\bibinfo{year}{2009}) \bibinfo{pages}{155431}.
\bibitem[{Ziambaras et~al.(2007)Ziambaras, Kleis, Schr\"oder, and
  Hyldgaard}]{Potassium}
\bibinfo{author}{E.~Ziambaras}, \bibinfo{author}{J.~Kleis},
  \bibinfo{author}{E.~Schr\"oder}, \bibinfo{author}{P.~Hyldgaard},
\newblock \bibinfo{title}{Potassium intercalation in graphite: A van der waals
  density-functional study},
\newblock \bibinfo{journal}{Phys. Rev. B} \bibinfo{volume}{76}
  (\bibinfo{year}{2007}) \bibinfo{pages}{155425}.
\bibitem[{Perdew et~al.(1996)Perdew, Burke, and Ernzerhof}]{PBE}
\bibinfo{author}{J.~P. Perdew}, \bibinfo{author}{K.~Burke},
  \bibinfo{author}{M.~Ernzerhof},
\newblock \bibinfo{title}{Generalized gradient approximation made simple},
\newblock \bibinfo{journal}{Phys. Rev. Lett.} \bibinfo{volume}{77}
  (\bibinfo{year}{1996}) \bibinfo{pages}{3865--3868}.
\bibitem[{DAC()}]{DACAPO}
\bibinfo{title}{Opensource code dacapo},
  \bibinfo{howpublished}{http://www.fysik.dtu.dk/CAMPOS/}, \bibinfo{year}{.}
\bibitem[{David et~al.(1991)David, Ibberson, Matthewman, Prassides, Dennis,
  Hare, Kroto, Taylor, and Walton}]{C60:nature}
\bibinfo{author}{W.~I.~F. David}, \bibinfo{author}{R.~M. Ibberson},
  \bibinfo{author}{J.~C. Matthewman}, \bibinfo{author}{K.~Prassides},
  \bibinfo{author}{T.~J.~S. Dennis}, \bibinfo{author}{J.~P. Hare},
  \bibinfo{author}{H.~W. Kroto}, \bibinfo{author}{R.~Taylor},
  \bibinfo{author}{D.~R.~M. Walton},
\newblock \bibinfo{title}{Crystal structure and bonding of ordered {C60}},
\newblock \bibinfo{journal}{Nature} \bibinfo{volume}{353}
  (\bibinfo{year}{1991}) \bibinfo{pages}{147--149}.
\bibitem[{{Londero} and {Schr\"oder}(2010)}]{VO5}
\bibinfo{author}{E.~{Londero}}, \bibinfo{author}{E.~{Schr\"oder}},
\newblock \bibinfo{title}{{Role of van der Waals bonding in layered oxide: Bulk
  vanadium pentoxide}},
\newblock \bibinfo{journal}{ArXiv e-prints}  (\bibinfo{year}{2010}).
\bibitem[{Lazi{\'c} et~al.(2010)Lazi{\'c}, Atodiresei, Alaei, Caciuc, Bl\"ugel,
  and Brako}]{JuNoLo}
\bibinfo{author}{P.~Lazi{\'c}}, \bibinfo{author}{N.~Atodiresei},
  \bibinfo{author}{M.~Alaei}, \bibinfo{author}{V.~Caciuc},
  \bibinfo{author}{S.~Bl\"ugel}, \bibinfo{author}{R.~Brako},
\newblock \bibinfo{title}{{JuNoLo - J\"ulich nonlocal code for parallel
  post-processing evaluation of vdW-DF correlation energy}},
\newblock \bibinfo{journal}{Computer Physics Communications}
  \bibinfo{volume}{181} (\bibinfo{year}{2010}) \bibinfo{pages}{371 -- 379}.
\bibitem[{Chakarova-K\"ack et~al.(2006)Chakarova-K\"ack, Schr\"oder, Lundqvist,
  and Langreth}]{Prl:svetla}
\bibinfo{author}{S.~D. Chakarova-K\"ack}, \bibinfo{author}{E.~Schr\"oder},
  \bibinfo{author}{B.~I. Lundqvist}, \bibinfo{author}{D.~C. Langreth},
\newblock \bibinfo{title}{Application of van der waals density functional to an
  extended system: Adsorption of benzene and naphthalene on graphite},
\newblock \bibinfo{journal}{Phys. Rev. Lett.} \bibinfo{volume}{96}
  (\bibinfo{year}{2006}) \bibinfo{pages}{146107}.
\bibitem[{Kleis(2006)}]{JesperPhD}
\bibinfo{author}{J.~Kleis}, \bibinfo{title}{Van der Waals density-functional
  description of polymers and other sparse materials}, Ph.D. thesis, Chalmers,
  \bibinfo{year}{2006}.
\bibitem[{Rom\'an-P\'erez and Soler(2009)}]{Soler:speedup}
\bibinfo{author}{G.~Rom\'an-P\'erez}, \bibinfo{author}{J.~M. Soler},
\newblock \bibinfo{title}{Efficient implementation of a van der waals density
  functional: Application to double-wall carbon nanotubes},
\newblock \bibinfo{journal}{Phys. Rev. Lett.} \bibinfo{volume}{103}
  (\bibinfo{year}{2009}) \bibinfo{pages}{096102}.
\bibitem[{Kleis et~al.(2008)Kleis, Schr\"oder, and Hyldgaard}]{Kleis:Nanotube}
\bibinfo{author}{J.~Kleis}, \bibinfo{author}{E.~Schr\"oder},
  \bibinfo{author}{P.~Hyldgaard},
\newblock \bibinfo{title}{Nature and strength of bonding in a crystal of
  semiconducting nanotubes: van der {W}aals density functional calculations and
  analytical results},
\newblock \bibinfo{journal}{Phys. Rev. B} \bibinfo{volume}{77}
  (\bibinfo{year}{2008}) \bibinfo{pages}{205422}.
\bibitem[{Kleis et~al.(2005)Kleis, Hyldgaard, and Schr\"oder}]{Kleis:polymer1}
\bibinfo{author}{J.~Kleis}, \bibinfo{author}{P.~Hyldgaard},
  \bibinfo{author}{E.~Schr\"oder},
\newblock \bibinfo{title}{{van der Waals interaction of parallel polymers and
  nanotubes}},
\newblock \bibinfo{journal}{Computational Materials Science}
  \bibinfo{volume}{33} (\bibinfo{year}{2005}) \bibinfo{pages}{192 -- 199}.
  \bibinfo{note}{Proceedings of the E-MRS 2004 Spring Meeting; Symposium H:
  Atomic Materials Design: Modelling and Characterization}.
\bibitem[{Bader(1990)}]{baderbook}
\bibinfo{author}{R.~F.~W. Bader}, \bibinfo{title}{Atoms in Molecules - A
  Quantum Theory}, \bibinfo{publisher}{Oxford University Press},
  \bibinfo{year}{1990}.
\bibitem[{Borck and Schr\"oder(2006)}]{obes}
\bibinfo{author}{{\O}.~Borck}, \bibinfo{author}{E.~Schr\"oder},
\newblock \bibinfo{title}{Adsorption of methanol and methoxy on the
  {$\alpha$-Cr$_2$ O$_3$(0001)} surface},
\newblock \bibinfo{journal}{Journal of Physics: Condensed Matter}
  \bibinfo{volume}{18} (\bibinfo{year}{2006}) \bibinfo{pages}{10751}.
\bibitem[{Borck()}]{Oyvind}
\bibinfo{author}{{\O}.~Borck}, \bibinfo{howpublished}{unpublished},
  \bibinfo{year}{.}
\bibitem[{Henkelman et~al.(2006)Henkelman, Arnaldsson, and
  J\'onsson}]{henkelman}
\bibinfo{author}{G.~Henkelman}, \bibinfo{author}{A.~Arnaldsson},
  \bibinfo{author}{H.~J\'onsson},
\newblock \bibinfo{title}{A fast and robust algorithm for bader decomposition
  of charge density},
\newblock \bibinfo{journal}{Computational Materials Science}
  \bibinfo{volume}{36} (\bibinfo{year}{2006}) \bibinfo{pages}{354 -- 360}.
\bibitem[{Becka and Cruickshank(1963)}]{Hexamine:Struc}
\bibinfo{author}{L.~N. Becka}, \bibinfo{author}{D.~W.~J. Cruickshank},
\newblock \bibinfo{title}{{The Crystal Structure of Hexamethylenetetramine. I.
  X-ray Studies at 298, 100 and 34 $^\circ$K}},
\newblock \bibinfo{journal}{Proceedings of the Royal Society of London. Series
  A. Mathematical and Physical Sciences} \bibinfo{volume}{273}
  (\bibinfo{year}{1963}) \bibinfo{pages}{435--454}.
\bibitem[{Arnautova et~al.(1996)Arnautova, Zakharova, Pivina, Smolenskii,
  Sukhachev, and Shcherbukhin}]{Hexamine:Cohesion}
\bibinfo{author}{E.~A. Arnautova}, \bibinfo{author}{M.~V. Zakharova},
  \bibinfo{author}{T.~S. Pivina}, \bibinfo{author}{E.~A. Smolenskii},
  \bibinfo{author}{D.~V. Sukhachev}, \bibinfo{author}{V.~V. Shcherbukhin},
\newblock \bibinfo{title}{Methods for calculating the enthalpies of sublimation
  of organic molecular crystals},
\newblock \bibinfo{journal}{Russian Chemical Bulletin} \bibinfo{volume}{45}
  (\bibinfo{year}{1996}) \bibinfo{pages}{2723--2732}.
\bibitem[{Gallucci et~al.(1986)Gallucci, Doecke, and Paquette}]{dod:struc}
\bibinfo{author}{J.~C. Gallucci}, \bibinfo{author}{C.~W. Doecke},
  \bibinfo{author}{L.~A. Paquette},
\newblock \bibinfo{title}{X-ray structure analysis of the pentagonal
  dodecahedrane hydrocarbon {(CH)20}},
\newblock \bibinfo{journal}{Journal of the American Chemical Society}
  \bibinfo{volume}{108} (\bibinfo{year}{1986}) \bibinfo{pages}{1343--1344}.
\bibitem[{Chickos and William E.~Acree(2002)}]{chemref}
\bibinfo{author}{J.~S. Chickos}, \bibinfo{author}{J.~William E.~Acree},
\newblock \bibinfo{title}{Enthalpies of sublimation of organic and
  organometallic compounds. 1910--2001},
\newblock \bibinfo{journal}{Journal of Physical and Chemical Reference Data}
  \bibinfo{volume}{31} (\bibinfo{year}{2002}) \bibinfo{pages}{537--698}.
\bibitem[{Baskin and Meyer(1955)}]{graphite:struc}
\bibinfo{author}{Y.~Baskin}, \bibinfo{author}{L.~Meyer},
\newblock \bibinfo{title}{Lattice constants of graphite at low temperatures},
\newblock \bibinfo{journal}{Phys. Rev.} \bibinfo{volume}{100}
  (\bibinfo{year}{1955}) \bibinfo{pages}{544}.
\bibitem[{Zacharia et~al.(2004)Zacharia, Ulbricht, and Hertel}]{coh:graphite}
\bibinfo{author}{R.~Zacharia}, \bibinfo{author}{H.~Ulbricht},
  \bibinfo{author}{T.~Hertel},
\newblock \bibinfo{title}{Interlayer cohesive energy of graphite from thermal
  desorption of polyaromatic hydrocarbons},
\newblock \bibinfo{journal}{Phys. Rev. B} \bibinfo{volume}{69}
  (\bibinfo{year}{2004}) \bibinfo{pages}{155406}.
\bibitem[{Zaremba and Kohn(1976)}]{ZarembaKohn}
\bibinfo{author}{E.~Zaremba}, \bibinfo{author}{W.~Kohn},
\newblock \bibinfo{title}{Van der waals interaction between an atom and a solid
  surface},
\newblock \bibinfo{journal}{Phys. Rev. B} \bibinfo{volume}{13}
  (\bibinfo{year}{1976}) \bibinfo{pages}{2270--2285}.

\end{thebibliography}
\end{document}